\documentclass[letterpaper]{article} 
\usepackage{aaai24}  
\usepackage{times}  
\usepackage{helvet}  
\usepackage{courier}  
\usepackage[hyphens]{url}  
\usepackage{graphicx} 
\usepackage{natbib}  
\usepackage{caption} 
\usepackage{algorithm}
\usepackage{algorithmic}

\usepackage{newfloat}
\usepackage{listings}
\DeclareCaptionStyle{ruled}{labelfont=normalfont,labelsep=colon,strut=off} 
\floatstyle{ruled}
\newfloat{listing}{tb}{lst}{}
\floatname{listing}{Listing}
\usepackage{tabularx}

\usepackage{multirow}

\usepackage[capitalize,noabbrev]{cleveref}

\title{An FDA for AI?  Pitfalls and Plausibility of Approval Regulation for Frontier Artificial Intelligence}
\author {
    Daniel Carpenter\textsuperscript{\rm 1},
    Carson Ezell\textsuperscript{\rm 1}
}
\affiliations {
    \textsuperscript{\rm 1}Harvard University\\
    dcarpenter@gov.harvard.edu, cezell@college.harvard.edu
}

\begin{document}

\maketitle

\begin{abstract}
Observers and practitioners of artificial intelligence (AI) have proposed an FDA-style licensing regime for the most advanced AI models, or 'frontier' models.  In this paper, we explore the applicability of \textit{approval regulation} -- that is, regulation of a product that combines experimental minima with government licensure conditioned partially or fully upon that experimentation -- to the regulation of frontier AI.  There are a number of reasons to believe that approval regulation, simplistically applied, would be inapposite for frontier AI risks.  Domains of weak fit include the difficulty of defining the regulated product, the presence of Knightian uncertainty or deep ambiguity about harms from AI, the potentially transmissible nature of risks, and distributed activities among actors involved in the AI lifecycle. We conclude by highlighting the role of policy learning and experimentation in regulatory development, describing how learning from other forms of AI regulation and improvements in evaluation and testing methods can help to overcome some of the challenges we identify. 
\end{abstract}

\section{Introduction}

Massive leaps in the scale, performance, and apparent risks of artificial intelligence (AI) have led practitioners and observers to call for various forms of regulation. 
Many proposals have focused on adapting AI regulations to account for the novel risks introduced by 'frontier' AI \cite{anderljung_frontier_2023, schuett_principles_2024}. Based on recent trends, we characterize frontier AI systems as usually having (1) general-purpose functionality, (2) more costly R\&D processes \cite{whittaker_steep_2021,ahmed_growing_2023}, (3) dual-use capabilities that pose misuse risks, and (4) systemic or structural risk \cite{zwetsloot_thinking_2019, council_of_the_european_union_proposal_2024}. 
Some regulatory proposals for frontier AI are analogs of regulatory regimes already in existence, including that of the U.S. Food and Drug Administration (FDA). 
For example, at a recent congressional hearing, emeritus professor Gary Marcus stated that among the ``many guardrails and regulations I would suggest,'' one was ``Creating an FDA-like regulatory regime for AI that evaluates large-scale deployment, balancing risks and benefit'' \cite{marcus_replies_2023}. There have been several proposals arguing for an explicit licensing regime based upon FDA-style approval regulation, including licensing that would apply to model development, deployment, or the operation of large datacenters \cite{stein_safe_2023, matheny_model_2023, microsoft_governing_2023, encode_justice_ai_2023, smith_licensing_2024, malgieri_licensing_2024, allen_roadmap_2024}.
Other aspects of FDA regulation have also served as analogies. For example, the National Artificial Intelligence Advisory Committee (NAIAC) has called for an adverse events reporting system, wherein the FDA system of the same title (it has been known for decades as the AERS) figures as a reference point \cite{the_national_artificial_intelligence_advisory_committee_naiac_recommendation_2023}. 

The possibility of an FDA-like regulatory regime for frontier AI has since occasioned considerable debate. AI-related regulatory frameworks and legislative proposals have involved various forms of licensing \cite{blumenthal_bipartisan_2023, center_for_ai_policy_model_2024}, and critics have identified many limitations of a licensing regime \cite{guha_ai_2023, wheeler_licensing_2024}, including a range of libertarian organizations and writers who quickly aligned against the idea \cite{bailey_openai_2023, thierer_problem_2023}. 
Thus far, the mapping of FDA-like analogies to AI regulation has largely proceeded by means of vague metaphors -- understandable for an early stage of public debate, but not desirable as actual policies are discussed. Indeed, there are properties of AI and its risks that would call for reconsideration of some aspects of FDA-style regulation as it has been traditionally practiced. 

In this paper, we argue that there is need for careful consideration of institutional and organizational forms before any regime, much less an ``FDA-like regulatory regime,'' could be adopted. We proceed through four general claims. First, at its essential core, FDA-style regulation is a form of \textit{approval regulation} linking mandatory testing with a regulatory veto over part or all of a firm's R\&D process. Second, this regime of regulation makes specific assumptions about the product and firm that are being regulated, the measurability of risks from the product, the observability of a firm's actions (e.g. development and testing), and the enforceability of rules that prevent certain unapproved activities from taking place. Third, there are aspects of frontier AI that do not conform to these assumptions. Finally, policy experimentation and learning are essential to addressing some limitations of approval regulation, including from other forms of AI regulation and developments in model evaluation methods. 

In Section 2, we describe approval regulation, including its properties that make it unique from other forms of regulation. In Section 3, we describe the conditions that facilitate approval regulation in the FDA context. In Section 4, we consider the applicability of approval regulation to the frontier AI context. In Section 5, we overview the role of policy experimentation and learning in the development an an approval regulation regime.

Before proceeding, two prefatory notes.  First, we make no judgment here about whether approval regulation is optimal or efficient in the spaces in which it has been applied, especially in the area of biomedical innovation.   
Second, it is important to consider the possible complementarity or substitutability of different regulatory policies.  Much of the argument from libertarian voices suggests that it is possible to rely upon self-regulation, intellectual property regulation, fiduciary or ``duty of care'' standards, or tort liability regimes to regulate AI harms \cite{thierer_flexible_2023}. These arguments may be on the mark, but it is worth noting that in many areas of regulation -- and not just biomedical innovation -- forms of approval regulation co-exist with these and other forms of governance.  To say that they co-exist is \textit{not} to assert that they do so without friction or inefficient cross-subsidization of activities. The point is that the desirability or plausibility of one form of regulatory institution does not, \textit{ipso facto}, rule out the possible desirability or plausibility of another. Considering the optimal portfolio of institutions is exactly where research is needed, and it is unlikely that any such portfolio will be designed \textit{ex nihilo} but will evolve.

\section{``FDA-Style''  
Approval Regulation}

Commentators referring to an ``FDA-like'' or ``FDA-style'' regime are usually referencing the FDA's regulation of new biomedical products, a function which is now global and exercised by dozens of national and regional regulators (the European Medicines Agency (EMA), for instance). At their core, ``FDA-style'' regimes rest upon structures of \textit{approval regulation} \cite{carpenter_protection_2004, carpenter_regulatory_2007, carpenter_approval_2010, henry_research_2019, henry_regulation_2022, ottaviani_approval_2023}, which we define here as a regime in which a regulator requires a firm to engage in testing before conducting subsequent activities (e.g. releasing a product), and in which this testing generates data that is used by the regulator to decide whether part or all of the subsequent activities can be conducted.  

So defined, approval regulation gives the regulator a ``veto'' over stages of product development and release, but approval regulation is far more than a mere gatekeeping function or a veto.  Any number of other regulatory mechanisms can regulate ``entry'' \cite{djankov_regulation_2002}, including mechanisms that are already being used in AI governance. For example, the EU AI Act requires ex-ante assessments of conformity with standards for high-risk AI deployments \cite{council_of_the_european_union_proposal_2024}.

Furthermore, many regulatory mechanisms can encourage or require testing, audits, and/or information disclosure without linkage to approval mechanisms. For example, Executive Order 14110 (EO 14110) in the U.S. requires that certain testing results for "dual-use foundation models" are reported to regulators \cite{executive_office_of_the_president_safe_2023}. Furthermore, various standards-setting bodies, including the National Institute of Standards and Technology (NIST) in the U.S. \cite{us_department_of_commerce_at_2023} and  CEN (European Committee for Standardization)/CENELEC (European Committee for Electrotechnical Standardization) in the E.U. \cite{laux_three_2024}, are developing further AI standards and best practices, including testing guidelines. Subsequent regulations can require or incentivize developers to follow these standards without a connection to approval mechanisms.

The essential properties of approval regulation were outlined in a series of mathematical models before 2010 \cite{carpenter_protection_2004, carpenter_regulatory_2007, carpenter_approval_2010}, and subsequent research has led to a more detailed understanding of how these regimes develop and operate. The history of approval regulation institutions has been the subject of studies in history and political science \cite{marks_progress_1997, carpenter_reputation_2010}. Models from the economic theory and management science literature have examine general properties of regulation and veto institutions and consider issues such as optimal timing of entry and regulation, the structure of costly experimentation in persuasion, and the relationship between \textit{ex ante} and \textit{ex post} regulation \cite{henry_research_2019, henry_regulation_2022, ottaviani_approval_2023}. Recent models have also explored a range of alternative institutional arrangements and the potential tradeoffs or complementarities among them \cite{henry_research_2019, mcclellan_experimentation_2022, bates_incentive-theoretic_2024}. While all of these models are simplifications, they are nonetheless essential for understanding the critical operative structure and incentive-based kernels of these regimes, especially when modelers pay appropriate attention to the institutional context. 

The \textit{combination} of testing and veto in approval regulation is essential to differentiating these institutions from other institutions that erect entry barriers. These two powers reinforce each other, create particular incentives, and complement a range of other regulatory policies implemented and enforced by agencies such as the FDA and EMA.  

The ability of regulators to write new rules governing testing depends heavily upon gatekeeping. Regulators verify that tests are conducted according to particular practices because specific testing results are necessary for them to perform their gatekeeping function.  
For example, required labeling for biomedical products incorporates information from required experiments, and the proposed labeling is an important part of the pre-market review. While regulators are influential in shaping and standardizing best practices in testing, their views of these best practices are also significantly influenced by developments that are exogenous to regulation. 
For example, the standards of pre-market review at the FDA developed hand-in-hand with changes in pharmacological and experimental standards \cite{marks_progress_1997, carpenter_reputation_2010}.  In terms of phased experimentation, developments in oncology (especially at the National Cancer Institute) were critical to the FDA's view of phased experiment \cite{carpenter_reputation_2010, keating_cancer_2014}.  

Approval regulation also creates particular experimental and long-range behavioral incentives. First, the fact that the regulator likely has a higher bar for converting R\&D into product launch than does the firm itself means that firms have incentives to conduct more testing than they otherwise would \cite{carpenter_regulatory_2007, henry_research_2019}, and adhere more closely to practices specified by the regulator. Notably, the primary costs associated with gatekeeping regulation are not the agency's decision itself but the set of experiments that come before, which are directly observed and regulated.
\footnote{In the model of \citet{carpenter_regulatory_2007}, the firm possesses a more precise prior on the state variable of the regulated product -- the asymmetric information is not absolute -- but all experiments are publicly observed.  Later approval regulation models have a similar structure, and while there are aspects of this assumption that are violated in the real world (such as when a regulatory sponsor has access to certain aspects of Phase III trial records that the regulator does not), this simplification captures much of the actual operation of approval regulation regimes.} Second, a single company likely has a range of products, some that are already released and others that are under development. A key property of the biomedical marketplace is that there is more profit to be made from the newest products than the older ones, due in part to patents \cite{carpenter_protection_2004, carpenter_early_2010}.  This means that even a profitable firm has strong incentives to behave ``well'' in front of the approval regulator, as its profitability depends heavily upon a stream of new molecules to be authorized in the future. Being perceived well by the regulator can both increase the chance of approval and reduce the expected time to approval.

Of course, the EMA and FDA do many things other than require testing and decide upon the marketability of new biomedical products.  These agencies inspect production facilities, require firms to conduct experiments after regulatory authorization, require firms and other actors to generate reports on ``adverse events'' associated with the product, monitor other data (a form of observational epidemiology), consider revisions to labels and warnings, and also regulate advertising and marketing practices.  How can we consider these in relation to approval regulation?  It is useful to differentiate here between the set of things that happen 
before a product is authorized -- \textit{ex ante} regulation -- and the set of things that happen afterwards -- \textit{ex post} regulation \cite{carpenter_reputation_2010, henry_regulation_2022}.  The basic structure of phased experiment -- Phase I trials for basic toxicity in non-diseased individuals, Phase II and III trials for examination of safety and efficacy in diseased populations -- occurs before marketing authorization (the ``veto'').  Yet important regulatory tools are available after regulatory marketing authorization. The regulator can require or request changes in labeling, can remove the product from the market (making the initial approval reversible at least in fact) and can, on its own volition, monitor a range of other data on the evolving risks of the approved product. 

\section{Conditions for Approval Regulation}

As it has developed in the area of biomedical innovation \cite{marks_progress_1997, carpenter_reputation_2010}, approval regulation assumes a particular form. A firm develops a molecule and then begins to test it, first upon non-human animals and then upon humans in a series of clinical trials.\footnote{Importantly, at the EMA and FDA, the relevant regulated organization (the ``firm'') is not necessarily the one that ``discovered'' the product (molecule) but its rather the ``sponsor,'' the firm that prepares and submits the regulatory dossier.  As detailed in \citet[Chapter~10]{carpenter_reputation_2010}, the structure of approval regulation at the FDA and related agencies is such that regulatory sponsorship is now an established, if not pivotal, component of biopharmaceutical firms.}  The regulator observes these trials and their results on roughly the same schedule -- though not, simultaneously, with the same precision -- as does the firm. The firm then collects data and documentation from these experiments and other tests (such as manufacturing data) and submits a ``new drug application'' or ``dossier'' to the regulator.  The dossier is massive and is the basis for the regulator's decision of whether or not to authorize/release for marketing of the drug.  After regulatory approval, the regulator often mandates further experiments (often called ``postmarketing trials'' or ``Phase IV trials'') and also monitors the risk profile of the molecule through a combination of inspections, adverse event reports and survey of databases. 
Medical device regulation carries forward many principles and institutions from drug regulation.  
In both molecules and devices, the dominant regulatory regimes for the FDA include mandatory pre-market experimentation and then an approval decision \textit{based upon those experiments}. 

The set of assumptions and enabling structures undergirding these regulatory regimes is considerable.  It includes:
\begin{itemize}
\item \textbf{Identifiability of a regulated unit}. In examining any regulatory policy, we should ask what is the thing to be regulated, to be governed? In the case of biopharmaceutical regulation, it is the molecule even more than the firm. More specifically and germanely, approval regulation in biopharmaceuticals generally possesses an identifiable object of regulation. This is not exogenous to regulation but is defined in part by the law itself, in the concepts of Investigational New Drug and New Molecular Entity or New Therapeutic Biological Products, or in the case of medical devices, Class III devices. 
\item \textbf{Identifiability of firms engaging in regulated activities}. In part because biomedical innovation is exogenously costly, in part because the costs associated with regulation itself, and in part because of the incentives stemming from patent systems (an agent must claim intellectual property rights over the molecule in order to enjoy patent protection upon its marketing authorization), the production of new therapeutic molecules and the agents or organizations that produce them and conduct experiments upon them is often well known. This assumption holds even in innovation markets with highly secondary and tertiary markets for contracting and sub-contracting. 
\item \textbf{Testing methods to identify and measure risk.} In biopharmaceutical regulation, two facts about the data used in evaluation are that (1) the adverse events to which probabilities are assigned are often known and detectible and (2) well-known probability models can be developed to describe the risk of these adverse events, such that these probability models are consulted directly in product evaluation. While in theory the set of things that could go wrong is infinite, in practice it is usually quite manageable.\footnote{This is even true with the transmissible risk from biologics, as in many cases infectious disease specialists know at least some, if not many, of the ``red flags" to look for.}  For instance, a vast amount of research has been conducted on the risk of hepatotoxicity associated with the ingestion of biopharmaceuticals, as many of these products place heavy demands upon the liver and their therapeutic properties often depend upon metabolization there.  An entire set of measurements and statistics are available for measuring these risks and assigning probabilities or severity measures to them. The ``set of things that could go wrong'' is often well known and regulators know where to look for most of (perhaps not all of) the risk.  Beyond this, the tests conducted by developers and required by regulators make it more likely that adverse events will be potentially observable at sufficient frequency that large-sample properties of statistical inference can be applied. 
\item \textbf{Observability of the fact of testing, once mandated}. In biopharmaceutical regulation, the event that ``the firm conducts a test upon its product'' is highly observable, and in models of approval regulation \cite{carpenter_protection_2004, carpenter_regulatory_2007, henry_research_2019}, this fact is perfectly observable and at a cost known to the regulator as well as the firm. This fact is in part endogenous to institutions, including regulatory institutions, because the molecule is registered with the FDA (all drugs under study in the United States must have an approved status of Investigational New Drug (IND)) as well as professional institutions (funding agencies such as the National Institute of Health, research clinics and hospitals that are regulated by professions and by numerous levels of government). Furthermore, groups of professional scientists and statisticians are routinely consulted in the design, pre-registration and analysis of these experiments. This observability assumes that regulators and experts external to the firm are given access to information about the product and the experiments conducted upon it.
\item \textbf{Observability of the fact of development}.  In biopharmaceutical regulation, it is difficult for actors to conceal the development, release, and marketing of new therapeutic products. For example, if a consumer wants insurance to pay for health services, they will need to come from a licensed or recognize provider, and relatedly, the product prescribed to the consumer will need to be listed on some kind of formulary.  In the market for human medical services as well as the market for therapeutic commodities (pharmaceuticals or devices), the vast insurance market serves as a \textit{de facto} regulator of illegal development and provision. It is not impossible, however, and substantial activity prevails at the margins of the regulated marketplace, either with known but unregulated products that are consumed (but not legally marketed) with believed health effects in mind, such as nutritional supplements, or with non-ethical drug use for health-related purposes (those who grow their own cannabis and who use it for self-ascribed health improvement reasons). In related forms of regulation, such as the regulation of new dams or nuclear reactors, the ability of an actor to ``innovate'' (create a new product) outside the bounds of regulation is again quite limited.  In the field of molecules, this fact is also not exogenous to institutions, as a range of drug enforcement agencies at various levels of government monitor and enforce laws against unauthorized production of chemical substances.
\item \textbf{An industrial structure and social institutions that facilitate the previous assumptions.}  The identifiability of firms, the ability of the regulator (or other agents) to observe these firms' behavior, and the observability of the fact of testing (a kind of compliance) are greatly facilitated in the biopharmaceutical industry by the fact that the number of firms, while large, is not so large as to defy manageability.  Once we consider the fact that the field for evaluating risk in biopharmaceuticals is often bounded by the extent of a diseased population, it is further the case that the number of firms and laboratories active in a particular disease market is far smaller than the set of all biopharma firms generally.  While there is no mathematical or empirical proof of the hypothesis, there may be reason to believe that the feasibility of approval regulation depends in part upon an oligopolistic industrial structure.  Beyond this, much of FDA governance in molecules and medical devices is assisted by, relies upon the science and professional standards of, and assumes the enforcement of physicians and other medical and health professions.

\end{itemize}

A final note.  Some observers might quibble, and fairly, with this simplified description of the biopharmaceutical world to which ``FDA-style'' approval regulation has been applied.  Our point is that these stylized facts have characterized something of the ``steady state'' of the biopharmaceutical world, even as it is an incredibly dynamic domain with massive amounts of investment and innovation.
Entire modes of innovation, from early forms of model-assisted drug development to the important role that AI itself now plays in drug development, have changed.  And yet some of the institutional and contextual features of the system are quite stable, and not only because of approval regulation.  

\section{Potential Pitfalls for AI Approval Regulation}

Given these stylized conditions that facilitate approval regulation, especially in the biopharmaceutical realm, we now turn to the emerging field of frontier AI development and assess the extent to which its characteristics are conducive to FDA-style regulation. Whether the facts adumbrated in the previous section apply to frontier AI regulation is an empirical question.  It is possible that the conditions for applicability of approval regulation to biopharmaceuticals, which are shown in \Cref{table:factors_table}, are not \textit{yet} satisfied in the area of frontier AI, but that they could be in the future, given policies or forms of industrial evolution, so nothing in this section should be construed as an impossibility result.  Another way of putting the matter is that \textit{the potential fit between models of approval regulation and frontier AI is a fruitful research agenda in institutional design as well as applied governance}. 

\begin{table*}[]
\centering
\begin{tabularx}{\textwidth}{|c|X|}
\hline
\textbf{Category} & \textbf{Considerations} \\
\hline
\multirow{4}{*}{Scope} & \textbf{Regulated Units:} What characteristics demarcate products or activities that are subject to approval regulation? \\ \cline{2-2} & \textbf{Regulated Entities:} How conducive are the organizational forms of entities involved in frontier AI development to facilitating oversight and complying with requirements? \\ \hline \multirow{4}{*}{Observability} & \textbf{Testing Requirements:} What evaluation tools and tests are available for measuring risks and informing approval decisions? \\ \cline{2-2} & \textbf{Oversight Mechanisms:} What oversight mechanisms are available for regulators to verify firms' compliance and ensure the rigor of model evaluation/testing or other forms of risk assessment? \\  \hline \multirow{3}{*}{Enforceability}  & \textbf{Control of Unregulated Activities:} To what extent do conditions enable unreported activities subject to regulation to persist, included unreported domestic activities or foreign activities that undermine the efficacy of domestic approval regulation? \\
\hline
\end{tabularx}
\caption{Conditions for the applicability of approval regulation to frontier AI, based on experiences from biomedical regulation.}
\label{table:factors_table}
\end{table*}

\subsection{Scope}

\subsubsection{Defining Regulated Units}
Approval gates are intended to regulate risky products or activities, so they rely on clear definitions of what counts as risky. However, frontier AI poses several challenges to such demarcations. 

First, there are not clear metrics to characterize the risk posed by AI systems, in part due to their generality and complexity. While definitions of foundation models that are dual-use or systemically risky have been used as the basis of regulatory action in the U.S. and E.U. \cite{executive_office_of_the_president_safe_2023, council_of_the_european_union_proposal_2024}, these definitions have been criticized for being overinclusive or underinclusive of certain types of systems that might be developed \cite{schuett_defining_2023, bommasani_drawing_2023}.

Furthermore, actors throughout the distribution chain can make a vast array of modifications to AI systems which can alter their risk profile \cite{davidson_ai_2023}, exacerbating the problem of defining what counts as a 'new' unit subject to gatekeeping. For example, fine-tuning or other modifications to parameters can alter a system's behavior or capabilities, and scaffolding frameworks in which AI systems are embedded can also alter their risk \cite{sharkey_causal_2024}. There is an open question of the extent to which this problem can be addressed by emerging methods that make it more difficult to modify models to introduce undesirable behaviors \cite{deng_sophon_2024, sheshadri_targeted_2024}. Another consideration is that a new foundation model with similar or identical properties to an existing model (e.g. architecture, data, etc.) might not count as a new system for the purpose of regulation.

The identifiability of homogeneous regulated units is not merely important for determining when a new system is developed that is subject to approval, but also for aggregating data about risk to enable more informed assessments \cite{bommasani_picking_2022}. If new data about an AI system (e.g. from testing or incident reports) leads to an updated risk profile, regulators rely upon an understanding of how applicable the new findings are to other systems. However, this task becomes more difficult as heterogeneity becomes more complicated.
The FDA has more established methods for data aggregation. For example, when examining a large dataset of chemical assays of a molecule, or the experience of thousands of patients with that molecule, or the mechanical properties of a hip implant, or the experiences of thousands of patients with said device, both the product and the experience have to sufficiently comparable (or ``commensurable'' as to be able to aggregated).

In addition, the development of a model itself introduces risks.
In biomedical innovation, there are many products and experiments that the public or regulators generally do not see or do not observe as thoroughly, and this is especially so for the products that ``fail'' in the sense of not having achieved market launch \cite{hwang_failure_2016}.  These products sit on the ``shelf" and there is not likely much of a risk of their being seized and deployed for other uses.\footnote{In some sense, intellectual property regimes address some of this risk, but in most regulated markets they address the risk of illegal appropriation for profit, not for misuse.} In the world of algorithms there seems little, beside strong intellectual property and cybersecurity protections, to prevent others from 
unsafely using them \cite{guha_ai_2023, nevo_securing_2023}, including through stealing model weights or developing a similar model of their own. This raises the question of whether approval should be required for certain forms of development activities prior to any deployment, which may include the conduct of a large training run, certain forms of modification after pre-training, or the operation of a large datacenter where regulated models are trained and stored.

\subsubsection{Regulated Entities}

The originators of foundation models are, for the moment and in general, well known. 
However, compared to a range of other regulated entities -- say bank holding companies regulated by the Federal Reserve and other national bank regulators, or biopharmaceutical and medical device companies as regulated by the FDA or EMA -- there is far less known about the industrial structure of the AI industry. This fact stems in part from the novelty of the industry and its rapid rise, but also from the fact of its non-regulation. Regulation often stipulates certain organizational forms be taken by a regulated organization (a compliance department, or a regulatory affairs department) that must then function as a liaison between the organization and the relevant regulatory agency.  These sub-organizations produce considerable data and fulfill reporting requirements. They function as a translator for the agency and make the regulated firm and its products more ``observable.'' 

It is unclear whether the industrial organization of foundation model development will lead to an industrial structure with these properties. There are reasons to think that the future of foundation model development will be characterized by high-cost research and development and by a small number of dominant firms whose models not only outcompete the models of other firms on a performance basis, but also learn about the strengths and weaknesses of those rival models and adapt. Indeed, the compute cost of training a frontier model is increasing rapidly \cite{whittaker_steep_2021, sevilla_compute_2022,cottier_trends_2023}.
As with many other capital-intensive industries, then, the number of operative firms would be reduced. Large and well-resourced companies or laboratories would more likely have the organizational and financial capacity to comply with intensive reporting requirements. At least OpenAI and Anthropic have established internal positions or teams responsible for documenting the implementation of catastrophic risk assessment practices for frontier models \cite{openai_preparedness_2023, anthropic_anthropics_2023}. 

However, several factors could enable many smaller organizations to be involved in the development of new foundation models, including reductions in the cost of development of frontier models from improvements in algorithmic efficiency \cite{ho_algorithmic_2024} or meaningful alterations via post-training enhancements \cite{qi_fine-tuning_2023}. These organizations are less likely to have the resources to engage in as rigorous compliance and reporting activities.

Another problem arises from the fact that the set of organizations that deploy foundation models may differ materially and appreciably from the set of labs that create them. If deployers are making consequential decisions that impact the risk profile, including implementing their own usage monitoring or other safety guardrails, regulators may have an interest in granting approval for actions by deployers rather than, or in addition to, upstream developers \cite{stein_safe_2023}. However, deployers may lack the ability to conduct as rigorous testing and reporting as upstream developers due to having less expertise, resources, and, perhaps most importantly, access to proprietary information about the system that is maintained by upstream developers \cite{bommasani_foundation_2023, hacker_regulating_2023,anderljung_towards_2023,casper_black-box_2024}. 

In addition, deployers would likely lack the institutional forms that facilitate observability and verification of compliance, especially where model weights are openly released \cite{seger_open-sourcing_2023, kapoor_societal_2024}. The general principle here is that approval regulation in the biomedical realm depends upon a set of social and economic institutions that developed alongside and somewhat separably from approval regulators like the FDA or EMA.  In the biomedical realm, the secondary market for the ``deployment'' of approved technologies is regulated by the professionalization of prescribers and, more implicitly but no less consequentially, by the tort system. Similar structures and institutional forms are still in their early stages for AI development, deployment, and usage \cite{solaiman_gradient_2023, eiras_risks_2024, gorwa_moderating_2024, shevlane_structured_2024}. Yet this raises the question for AI regulation of what social and economics structures -- professionals that regulate use, tort systems that impose liability constraints, concentrated industrial structure that enhances the prospect for compliance capacity -- will emerge in foundation models.

\subsection{Observability}

\subsubsection{Experimentation Requirements}

In an important observation, \citet{knight_risk_1921} described a form of ``uncertainty'' in which events can be enumerated but probabilities cannot be assigned to them. In a recent paper, \citet{sunstein_knightian_2023} reviews the postulates of this concept and argues that regulatory policy development must take account of this ineluctable fact.

Whether probabilities can be assigned to the various risk events that we encounter with the development of AI to form the basis of effective regulatory decisions is not known. The complexity of the deployment environment means that model behaviors and their resulting effects are difficult to anticipate \cite{weidinger_sociotechnical_2023}. 

But even if Knightian uncertainty did not exist in this world, another problem would: deep ambiguity or what \citet{kay_radical_2020} call ``radical uncertainty.''  Compared to most regulated domains, the AI domain seems replete with potential risks and rewards that are, almost by forcible extension from the promise and pitfalls of artificial intelligence, hard to imagine. A related concern is what \citet{taleb_antifragile_2014} has called \textit{the Lucretius problem}, namely the tendency to believe that the past contains the full set of harms that could occur and that nothing worse than what is in that (memory) set could possibly occur in the future. 
This problem is exacerbated by increasingly capable models that can create previously unknown pathways for risk 
\cite{shevlane_model_2023, openai_preparedness_2023}. Or the auxiliary risks from diffusive bioweapons, proliferating nuclear weapons, or interconnectedness may exacerbate the harm that could happen from an otherwise stable process governing risks from foundation models \cite{zwetsloot_thinking_2019}. This makes risk evaluation and risk management not merely a difficult proposition but also requires those who would regulate frontier AI to consider scenarios that have never before occurred \textit{and have not yet been imagined}, either by machine or by human. 

Recent regulatory developments suggest that one kind of testing that is and will be conducted upon foundation models is 'red teaming' \cite{ganguli_red_2022,perez_red_2022, rando_red-teaming_2022,casper_explore_2023, feffer_red-teaming_2024}, which EO 14110 defines as a "structured testing effort" that usually involves "adversarial methods to identify flaws and vulnerabilities, such as harmful or discriminatory
outputs from an AI system, unforeseen or undesirable system behaviors, limitations, or potential
risks associated with the misuse of the system" \cite{executive_office_of_the_president_safe_2023}. The mapping from AI red teaming to risk assessment is akin to financial stress-testing, where red teaming exercises can provide insights into harms that can arise in various scenarios or contexts. Indeed, stress testing in the financial sector involves considering worst-case scenarios and contexts involving systemic risk. While previous AI red-teaming efforts have been limited \cite{feffer_red-teaming_2024}, more rigorous red-teaming exercises may involve providing red-teamers with greater access to models to assess worst-case behaviors \cite{kinniment_evaluating_2023,casper_black-box_2024}. Advances in methods to identify a wider array of undesirable behaviors can enhance the effectiveness of red-teaming. However, red-teaming shares a property of stress testing that tests are biased towards studying scenarios that humans have already imagined. Indeed, stress tests were conducted before the 2008 financial crisis, but they did not imagine and test for a scenario with sufficient stress from a decline in housing prices \cite{frame_failure_2015}. 

Behavioral testing methods, such as AI red-teaming as it is often practiced, are also prone to producing misleading results \cite{casper_black-box_2024}, including due to a poor understanding of training dynamics \cite{schaeffer_are_2023} or data contamination \cite{golchin_time_2023, oren_proving_2023}. A related concern is that the training process might encourage advanced models to behave well during behavioral testing in contrast to actual deployment \cite{berglund_taken_2023,cohen_regulating_2024, ngo_alignment_2024, hubinger_sleeper_2024}.  

Testing can involve complementary approaches beyond behavioral evaluations. For example, interpretability methods focused on studying model internals can gain insights into model reasoning \cite{wang_interpretability_2022, li_emergent_2022}. While researchers often use these methods to analyze small models \cite{elhage_toy_2022}, recent work raises the question of whether emerging interpretability methods can produce valuable insights about the behavior of large models \cite{cunningham_sparse_2023, marks_sparse_2024,templeton2024scaling}. "In-the-wild" testing \cite{naihin_testing_2023}, including in sandbox environments \cite{park_generative_2023}, can also produce insights that are difficult to produce in more controlled settings. Ecosystem-wide documentation \cite{bommasani_ecosystem_2023, chan_visibility_2024} can also be directly consulted to inform risk assessment. In general, various forms of testing can complement each other when assessing the risk profile of a model.

The question of risk from frontier AI is not merely the question of considering various pathways and their likelihoods, but also \textit{the potential costs incurred once that barrier is ruptured} (or ruptured with sufficient severity that serious human costs occur). To be clear, any regulatory regime that makes decisions based in part on imagined worst-case scenarios 
would have to avoid implementing the most na\"{i}ve decision rules.  Just because an exercise can produce a horrific imagined result -- the end of the world -- should not imply that the most restrictive regulatory response should be adopted \cite{sunstein_worst-case_2009}. Any speculative exercise that included the worst possible scenario would also need to consider humanity's likely best response in addition to regulatory options.

Despite ongoing experimentation and recent progress in AI risk assessment, the risks are currently far better known in biomedical innovation, in part because they have been known descriptively for decades or even a century or more.  We can and do measure the risk of liver damage or hepatotoxicity from drugs, but beyond that, there is abundant community knowledge about where such risks can lead and the likely profile of costs that can be imposed.  In oncology, for instance, there is an entire subfield dedicated to studying the cardiac risks of oncologic therapies, including cytotoxic and immunotherapeutic interventions \cite{herrmann_defining_2022,lyon_baseline_2020}.  The ``event'' (hepatotoxicity, cardiotoxicity) can be defined, as can its attendant sequellae that imposes costs upon the human person (conditional probability or likelihood of dysfunction requiring a transplant, or mortality).  Or in disaster insurance, there are entire industries dedicated to modeling the aggregate effects of a hurricane or tornado cluster. In short, there are a set of questions that any implementable risk science would need to be addressed in any risk-benefit analysis of a foundation model.

\subsubsection{Oversight Mechanisms}

If tests are required, what is the enforcement regime for ensuring that they are carried out?  Even in the area of biomedical regulation, many pivotal trials are not reported and many post-approval trials are neither commenced, completed nor fully reported \cite{carpenter_reputation_2010,moore_development_2014,hwang_postmarketing_2014,wallach_postmarket_2018}.  One descriptive study of new drugs approved by the FDA in 2008 found that five years later (2013)  ``26 of 85 (31\%) of the postmarketing study commitments had been fulfilled, and 8 (9\%) [of those studies] had been submitted for agency review'' 
\citetext{\citealt{moore_development_2014}; see also \citealt{carpenter_can_2014}}.   

\begin{figure*}[t!]
    \centering
    \includegraphics[width=.8\linewidth]{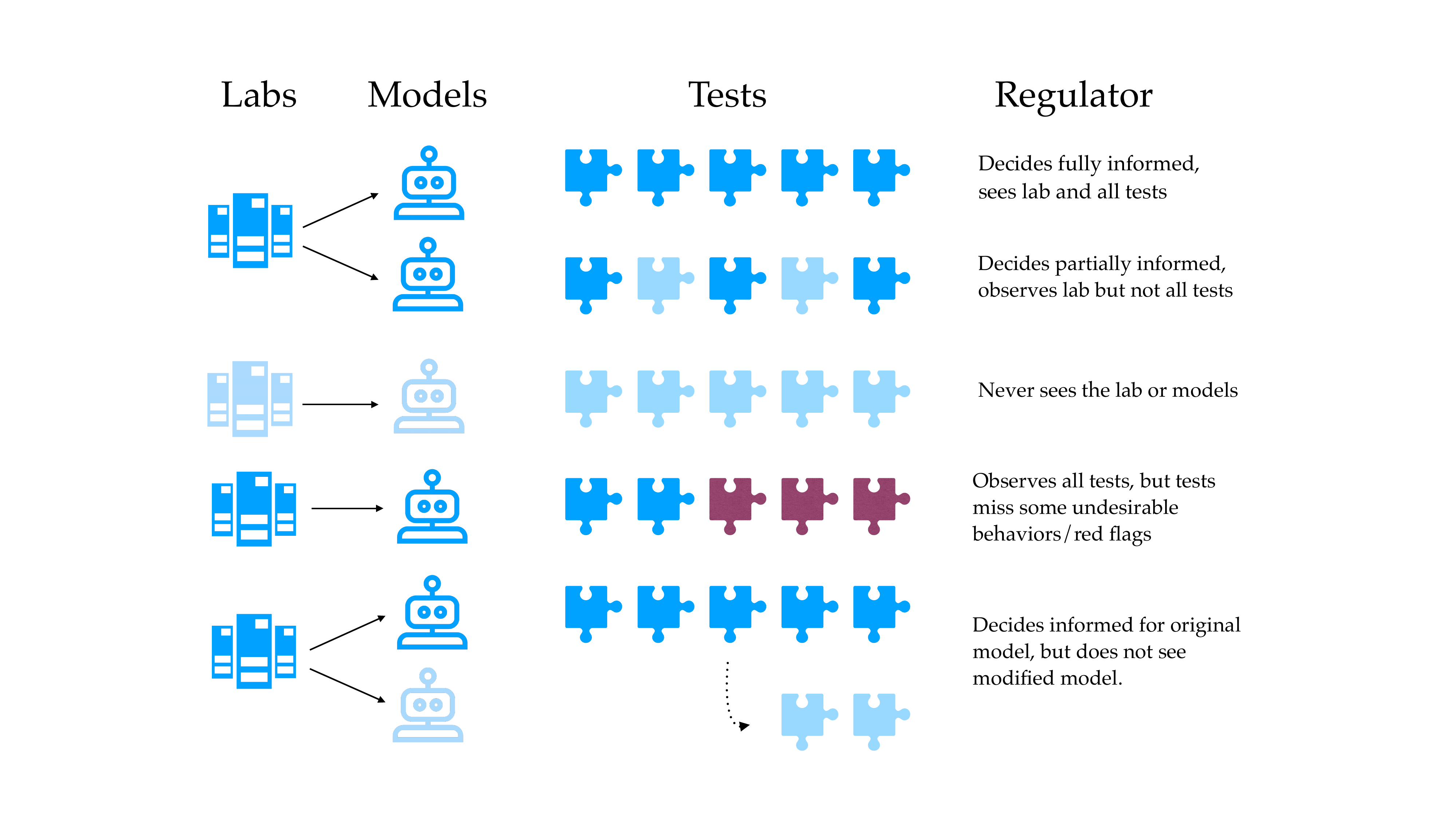}
    \caption{Possible scenarios where regulators lack complete information about frontier AI model development or testing. Arrows show which models were trained by which labs. Dark blue icons reflect regulators having complete information, and faded blue icons reflect a lack of regulatory visibility. Purple icons represent misleading or uninformative test results.}
    \label{fig:figure1}
\end{figure*}

The regulated organization would be responsible for carrying out testing or permitting government or third-party observers access to the resources with which they could be performed. In the case of financial stress tests, the regulated organization is often one of the most heavily regulated and well-documented organizations on the planet.  Consider, for example, the kinds of data that the Federal Reserve carries and published on commercial banks or bank holding companies (\url{https://www.federalreserve.gov/data.htm}).  On a quarterly basis, regulators observe hundreds if not thousands of indicators on the operation of each entity they regulate.  In the case of bank holding companies, for instance, this incudes a regular statement of their consulting, advising and external legal expenses \cite{libgober_lawyers_2024}.  And as of May 2022, different government agencies employ over 60,000 bank examiners.\footnote{See the data adduced by the Bureau of Labor Statistics, which decomposes the bank examiner population into several professional types; \url{https://www.bls.gov/oes/current/oes132061.htm}.}  

A similar degree of oversight does not exist for frontier AI developers. Many developers are hesitant to share information about their proprietary models \cite{bommasani_foundation_2023} and have incentives to limit information sharing \cite{casper_black-box_2024,kolt_responsible_2024}. Furthermore, documentation and disclosure practices among frontier model developers are inconsistent and incomplete \cite{kolt_responsible_2024, pal_model_2024}. This inhibits an understanding of model behaviors and risks even when information is shared, and some relevant information (e.g. detailed documentation of internal testing) might not be documented at all. In \Cref{fig:figure1}, we show several scenarios which could result in regulators lacking complete information about models subject to approval regulation, including testing data.

\subsection{Enforceability}

\subsubsection{Control of Unregulated Activities}

Approval regulation and any other kind of licensing or entry regulation depends upon institutions of detection. Furthermore, approval regulation is reinforced by the existence of an organization that could be sanctioned for illegally marketing or distributing an unapproved product, or that could potentially be fined for failure to observe regulatory requirements.

Direct regulation of innovators is becoming a standard feature of policy proposals in the AI domain.  This is the direction in which the Biden Administration in the United States \cite{executive_office_of_the_president_safe_2023} as well as the European Union are moving. The question becomes how enforceable such requirements are. The applicability of approval regulation to frontier AI governance depends, again, upon the existence, whether designed or co-evolved, of an industry structure that permits detection of R\&D, violation of regulatory requirements, and feasible compliance activities.

If firms or labs do not wish to announce the development of a new model, or if there are many small labs capable of producing new foundation models, then it may be more difficult for any third-party agent to observe many acts of a new foundation model being developed or deployed. 

However, foundation model development that is not reported by developers may be observable to compute providers because of the scale of frontier AI model training runs \cite{sastry_computing_2024}. 
As new AI models are developed and deployed, they often require massive utilization of computing power (and, relatedly, monetary investments to purchase relevant equipment, processing time and concomitant utilization of energy), so they are trained in large datacenters \cite{pilz_compute_2023}.  If these expenditures can be measured by regulators or third parties,  
then development of new foundation models may be detectible \cite{shavit_what_2023, heim_governing_2024}.  Another possibility is that the expense of new model development may be so high as to induce exogenous barriers to entry and a small number of dominant firms or labs.  Then as with the earlier problem of regulated organizations, industrial structure -- something like an oligopoly -- may reduce the set of regulable players to a manageable number. 

Still, some risk may come from the fact that organizations in less well-regulated jurisdictions may wish to invest in laboratories or compute infrastructure to develop their own AI capabilities -- this may include state-sponsored organizations. In addition, regulatory settings with lower-cost regulatory requirements might well attract more laboratory activity from abroad and compute infrastructure development. Because foundation models present the prospect of highly diffusive and contagious risk, the globalization or ``harmonization'' of regulatory requirements would likely be far more important in regulating AI than it would in regulating biomedical products \cite{ho_international_2023, trager_international_2023}. Hence, an effective approval regulation regime would still be constrained without
legal uniformity (at least for regulatory minima) for foundational model developers \cite{cihon_standards_2019}, as well as international coordination on monitoring development \cite{trager_international_2023,shavit_what_2023}.

\section{Developing Approval Regulation through Experimentation and Learning}

The upshot of these considerations might be that the conditions do not yet exist for the implementation of an effective approval regulation regime. However, it would be premature to conclude that the obstacles we laid out could not be overcome, especially through policy experimentation and learning. Any regulatory policy must be considered in a dynamic context, which means that \textit{the status quo must always be regarded as at least partially an experiment from which lessons can be drawn and to which adaptations can be made}. The longer history of approval regulation in molecules has taken the better part of a century (in devices, a half-century at least) to evolve, and decades- or century-long time horizons have characterized the evolution of regulation in other domains such as antitrust, anti-collusion, consumer product safety and systemic finance. 

Consider that molecular regulation in therapeutics started without a regulatory veto for therapeutic drugs—the 1906 Pure Food and Drugs Act gave the federal government post-market inspection and product removal power (though note that the very first vaccines \textit{did} have something like a gatekeeping institution in the 1902 Biologics and Vaccines Act).
Regulatory development depended heavily upon coincident developments in pharmacology, statistics and the study of clinical trials and cancer therapeutics \cite{keating_cancer_2014}. It was these developments, combined with particular regulatory crises, that led to a new regime of regulatory pre-market review in the 1930s and the subsequent Kefauver-Harris Amendments of 1962 \citep[Chapter~3]{carpenter_reputation_2010}, which mandated proof of "effectiveness".

Furthermore, regulatory experience at the FDA spurred scientific findings that have led to various transformations of its approval models. For example, in some areas of therapeutics, while there is an abiding debate about the merits of such programs \citep{fleming_surrogate_2005, moore_development_2014,carpenter_can_2014, budish_firms_2015, naci_characteristics_2017}, most or all new drugs are now approved on the basis of surrogate endpoints \cite{yu_use_2015}. 
The basic idea is that what society most cares about is mortality and morbidity, but that stand-in correlates of these core variables (tumor growth in solid tumors, say, or A1C reduction in diabetes medications) can be observed or measured earlier in the experimentation process, and may be sufficient for making decisions about the marketability of a new product.

\subsection{Learning from Other AI Regulations}

One possibility, and a scenario that has some historical experience to support its plausibility, is that "lighter" and more inchoate forms of regulation may generate lessons applicable to regulatory reform.  A range of governance proposals and regimes have already emerged for AI and foundation models. Regulation of foundation models is trending toward the adoption of registration and reporting requirements, and there are many aspects of these regimes, too, that suffer from adaptability and feasibility problems \cite{guha_ai_2023}. Several of the challenges are similar and may offer applicable lessons, including establishing definitions of regulated systems \cite{schuett_defining_2023,bommasani_drawing_2023}, conducting informative tests, establishing mechanisms for oversight of development, and limiting non-compliant activities. In some sense, there is relevant experimentation right now. 

In addition, early forms of foundation model regulation can precipitate the development of similar regimes in other jurisdictions. Such patterns have already emerged with institutions focused at least in part on frontier AI governance—the establishment of the UK AI Safety Institute \cite{department_for_science_innovation__technology_introducing_2024} was followed by the establishment of similar institutions in at least the U.S. \cite{us_department_of_commerce_at_2023}, Japan \cite{shimbun_japan_2023}, and Canada \cite{cass-beggs_welcome_2024}. Indeed, with pharmaceutical regulation, there have been adaptations of FDA-like regulatory frameworks adopted across national and regional settings. These patterns can increase international coordination and facilitate more experimentation, although more experimentation across national contexts trades off against harmonization and creates risks of regulatory 'races to the bottom'. 

European societies were long accustomed to apply less stringent approval regulation to pharmaceuticals than in the United States. The reduced stringency took several forms: (1) weaker experimental standards entailing less costly experiments that observed fewer dimensions of efficacy and risk, (2) weaker requirements on dossiers such that experimental data were summarized and not fully reported, and, finally, (3) easier approval standards. Counter-intuitively from the perspective of regulatory ``races to the bottom,'' it is Europe that moved in the direction of the United States, not vice versa \citep[Chapter~12]{carpenter_reputation_2010}. Many observers now consider European biopharmaceutical regulation to be more stringent than in the United States.

\subsection{Learning from AI Evaluation and Testing}

Methods for model evaluation and testing are being developed and iterated upon both in the context of emerging regulatory regimes, as well as within the AI research community more broadly \cite{chang_survey_2024, birhane_ai_2024}. These developments include novel benchmarks \cite{srivastava_beyond_2023,li_wmdp_2024}, methodologies and tools \cite{ kinniment_evaluating_2023, ojewale_towards_2024, hubinger_sleeper_2024}, taxonomies of risks \cite{weidinger_sociotechnical_2023, openai_preparedness_2023, critch_tasra_2023, shevlane_model_2023, anthropic_anthropics_2023}, and documentation practices for communicating results \cite{gilbert_reward_2023, kolt_responsible_2024, clymer_safety_2024}. Thus far, evaluation practices for frontier models have been largely unstandardized \cite{feffer_red-teaming_2024}, but they have produced key learnings \cite{ganguli_challenges_2023}, and there are nascent efforts to increase standardization of evaluation practices \cite{metr_portable_2024, us_department_of_commerce_us_2024, ai_safety_institute_inspect_2024}. 

One possibility is that a set of potentially governable risks and tools to measure them might be adduced as they emerge in either experimentation or in real-world behavior.  This is in the spirit of reporting requirements and incident reporting systems \cite{mcgregor_preventing_2021, the_national_artificial_intelligence_advisory_committee_naiac_recommendation_2023}.  The many decades of experience with adverse event reporting systems in biomedical innovation suggest that it will take considerable time and institutional investment to develop standardized frameworks for evaluation.

The scope of this nascent evaluation and risk detection industry is beyond the ambit of this paper.  An important question for those proposing approval regulation regimes \cite{stein_safe_2023}, a variety of ``FDA-like'' institutions \cite{tutt_fda_2016} or even ``adverse event reporting systems'' \cite{the_national_artificial_intelligence_advisory_committee_naiac_recommendation_2023}, however, is whether a standardized framework for threat detection and risk evaluation can emerge from these scattered efforts.  One may wish for a less standardized approach, but a true ``system''-based approach to regulation will, sooner or later, seek to aggregate across different datasets and analyses.\footnote{Another way of putting the question here is whether any unified regulatory regime should exist at all, as opposed to a range of less centralized arrangements operating in communication, but not stringent coordination, with each other. This is quite different from calls for self-regulation or no regulation at all.}  
In the biomedical regulation world, there have been decades of calls for ``harmonization'' of regulatory requirements and standards across nations.  The prima facie logic inspiring these proposals seems defensible, but given that federalism is itself a form of experimentation \cite{volden_formal_2008, callander_experimentation_2015}, one worries that learning value is surrendered when regulatory harmonization develops into strong uniformity.

One final problem with an experimental and incremental approach to regulation 
is that the materialization of the most severe risks may create conditions from which it is hard to escape. The most ``catastrophic'' risks from foundation model development may call for more stringent regulation in the first place \cite{stein_safe_2023, weil_tort_2024, cohen_regulating_2024}.

\section{Conclusion} 
This paper joins other calls for circumspection in the application of regulatory models to generative artificial intelligence, in particular calling for more careful consideration of the feasibility of ``FDA-like'' approval regulation regimes to the regulation of frontier AI models and the catastrophic risks they may pose.  The greatest impediments to such a model, in our judgment, are (1) enforceability of rigorous testing requirements and development/deployment restrictions and, perhaps most important, (2) the inapposite mapping between AI evaluation and the world of large samples and well-defined risks in which approval regulation operates, due to the lack of well-established indicators of catastrophic risk. However, we propose viewing these obstacles through the lens of policy learning, where the emergence of a regulatory regime that achieves fit within its domain depends upon adaptation and incorporation of new information from both regulatory experience and exogenous factors. 

Regulatory change, of course, implies neither regulatory evolution in a ``fitness'' sense nor monotonic improvement.  Yet in a range of domains, it is at least plausible that regulation has been transformed due to criticism, scientific analysis, benefit-cost analysis and more rational forms of political oversight \cite{mccraw_prophets_1986}.  This may not rise to the level of the culture championed by \citet{greenstone_toward_2009}, but that does not mean that useful information cannot be yielded by such learning, nor does it mean that a less formally experimental approach is worse.  Learning about policies from prospectively designed experiments alone may be difficult over the long run, and recent arguments suggest that a purely experimental approach may be wrong for optimization of policies in different domains \cite{stevenson_cause_2023}. Whatever the preferred mode of policy learning, it would be essential to approach such inferences prospectively and retrospectively, and to consider hybrid forms of regulation, given the rapidly changing nature of foundation models in AI and the often unquantifiable nature of their dangers.

Our point is not that approval regulation is a necessary component or end point of a comprehensive regulatory regime, or that other forms of regulation are necessarily insufficient or instrumental. There is, of course, no law that stipulates (and certainly no evidence consistent with any law that suggests) that regulation evolves in any monotonic fashion from less to more efficient.  Yet regulatory reform and deregulation have occurred in many domains \cite{greenstone_toward_2009}.  There is no unidirectionality to regulation.  Nor is there any systematic historical or empirical evidence for any such unidirectionality.

\section*{Acknowledgements}
The authors would like to acknowledge Open Philanthropy for supporting this project. In addition, the authors are grateful for discussions and feedback from the following individuals: Markus Anderljung, Sam Bell, Tyler Cowen, Lindsey Gailmard, Holden Karnofsky, Aaron Kesselheim, Trevor Levin, Michael Sklar, Kevin Wei, and Glen Weyl.

\bibliography{aaai_references.bib}

\end{document}